# Superionic-like diffusion in an elemental crystal: bcc Titanium

D.G. Sangiovanni,[1,2*] J. Klarbring,[1] D. Smirnova,[2,3] N.V. Skripnyak,[1] D. Gambino,[1] M. Mrovec,[2] S.I. Simak,[1] I.A. Abrikosov[1,4]

[1]Department of Physics, Chemistry and Biology (IFM) Linköping University,
SE-581 83, Linköping, Sweden

[2]ICAMS, Ruhr-Universität Bochum, D-44780 Bochum, Germany

[3]Joint Institute for High Temperatures, Russian Academy of Sciences, Moscow, Russia

[4]Materials Modeling and Development Laboratory, NUST "MISIS", 119049 Moscow, Russia

Recent theoretical investigations [Belonoshko *et al.* Nature Geoscience **10**, 312 (2017)] revealed occurrence of concerted migration of several atoms in bcc Fe at inner-core temperatures and pressures. Here, we combine first-principles and semi-empirical atomistic simulations to show that a diffusion mechanism analogous to the one predicted for bcc iron at extreme conditions is also operative and of relevance for the high-temperature bcc phase of pure Ti at ambient pressure. The mechanism entails a rapid collective movement of numerous (from two to dozens) neighbors along tangled closed-loop paths in defect-free crystal regions. We argue that this phenomenon closely resembles the diffusion behavior of superionics and liquid metals. Furthermore, we suggest that concerted migration is the atomistic manifestation of vanishingly small ω-mode phonon frequencies previously detected via neutron scattering and the mechanism underlying anomalously large and markedly non-Arrhenius self-diffusivities characteristic of bcc Ti.

*Corresponding author: davide.sangiovanni@liu.se  davide.sangiovanni@rub.se

Mass transport in elemental crystals is primarily regulated by migration of intrinsic point defects such as vacancies and self-interstitials. Vacancies and self-interstitials, present at dilute concentrations at equilibrium, are generated/annihilated at, e.g., surfaces and grain boundaries, or via Frenkel-pair formation/recombination in defect-free lattice regions. Diffusion of individual point-defects is often well described by uncorrelated hops among vicinal stable positions, which allows evaluation of diffusivities via simple stochastic-walk models. However, the occurrence of more complex migration mechanisms in elemental crystals has also been proposed. For example, experimental and theoretical studies indicated that self-interstitials propagate extremely rapidly in bcc metals [1,2] due to the long-range compressive strain that they produce in <111> lattice rows (defects known as crowdions [3-5]). Other diffusion mechanisms proposed for pure metals include spontaneous concerted exchange of two, or more, atoms in a defect-free lattice. In particular, in 1950, theoretical results by Zener indicated that synchronized cyclic motions of four atoms in fcc copper (4-atom ring mechanism) is energetically more favored than direct exchange of two neighbors [6].

Among elemental crystals, Group-IVB bcc transition-metals have attracted significant interest due to large atomic self-diffusivities $D$ and sharply non-Arrhenius trends in $D$ vs. inverse temperature [7]. The mechanism(s) that control diffusion in Group-IVB bcc metals have long been under debate [8]. Hypothesis made to explain the anomalous temperature dependence of self-diffusion coefficients include mixture of diffusion mechanisms and temperature-induced modifications of the effective activation energy for a single migration pathway [9]. In this regard, tracer-diffusivity experiments [10] excluded the possibility of 4-atom ring diffusion (proposed in [6]) controlling mass transport in bcc Ti and Ti–Nb alloys. Nonetheless, the authors [10] asserted that the occurrence of diffusion reactions involving concerted motion of ≥8 Ti neighbors would be consistent with their measurements.

Complex diffusion mechanisms are known for exotic type of materials such as superionic conductors [11-14], superheated crystals [15], and supercooled liquids [14,16,17]. While specific details of the mechanism in these cases differ, they all entail distinctively concerted migration of several atoms, often referred to as string-like or liquid-like diffusion. Recent theoretical investigations by Belonoshko *et al.* [18] identified concerted migration in bcc Fe subject to core temperatures and pressures. The results of [18] are intriguing as they suggest that rapid concerted diffusion may be a phenomenon of relevance even in elemental crystals. Nevertheless, experimental observation of liquid-like atomic migration in bcc Fe is practically unfeasible due to the extreme conditions necessary to dynamically stabilize this metallic phase [18]. Here, we combine classical and *ab initio* molecular dynamics (CMD, AIMD) to demonstrate that highly collective diffusion processes are active at temperatures safely below melting in the bcc phase of pure Ti at ambient pressure. We show that the occurrence of concerted migration, i.e. simultaneous diffusion of ≥2 atoms in defect-free lattice regions, may resolve longstanding questions concerning mass transport and phonon properties of Group-IVB bcc metals.

AIMD [19] simulations are carried out using VASP [20] implemented with the projector augmented-wave method [21], employing the Perdew-Burke-Ernzerhof [22] approximation to the electronic exchange/correlation energies. The ionic positions are updated every fs, using a convergence criterion of $10^{-5}$ eV/supercell for the self-consistent electron density, 2×2×2 k-point grids and planewave cutoff energies of 300 eV. AIMD simulations of defect-free bcc Ti (250 atoms) sample the canonical *NVT* configuration space at 1800 K, which is well above the range of bcc→hcp and bcc→ω transition temperatures (≈800–1100 K) [23], while below the Ti melting point $T_m$≈1940 K. The temperature is controlled via the Nosé-Hoover thermostat, using a Nosé mass of 148 fs. Supercell equilibrium volumes are taken from experiments [24].

CMD simulations are performed using LAMMPS [25], describing the Ti-Ti interactions via the second-nearest-neighbor modified embedded-atom method (MEAM) [26], as parameterized in [27], and integrating the equations of motion at 1-fs timesteps. The equilibrium volumes of supercells formed of 14×14×14 conventional bcc cells (5488 Ti atoms) are determined as a function of temperature via *NPT* sampling based on the Parrinello-Rahman barostat [28] and the Langevin thermostat. Concerted migration rates $\Gamma_C$, diffusivities $D_C$ in a defect-free lattice, and vacancy diffusivities $D_V$ are separately determined by combining the results obtained for Ti supercells comprised of 5488 and 432 bcc sites, respectively, during *NVT* canonical sampling at temperatures within the range $0.73–0.97 \cdot T_m$ [29]. Details on the evaluation of $D_V$, $\Gamma_C$, and $D_C$ with corresponding activation energies $E_a$, attempt frequencies $\nu$, and prefactors $D_0$ are given in [30]. The MEAM potential, carefully parameterized via matching the model forces with AIMD forces [27], correctly reproduces the mechanical instability of bcc Ti at low temperatures, as well as hcp→bcc transformations, but is unable to describe the phonon instability which drives spontaneous bcc→ω transitions at 0 kelvin. This, however, is not expected to qualitatively affect the conclusions of this work, which are based on results obtained at temperatures much higher than bcc→ω transition temperatures.

To estimate equilibrium vacancy and self-interstitial concentrations ($c_V$, $c_{SI}$), and total self-diffusion coefficients $D_{tot}$ [31] as a function of temperature *T*, we employ the approach proposed in [32] for a (initially defect-free) bcc Ti(001) surface slab (165600-atom supercell). CMD simulations show that point defects in the slab interior are generated by thermal fluctuations (Frenkel pairs) or had originally formed at the surface. The open surfaces allow maintaining the equilibrium balance between different defects [30].

In addition to the validation procedure performed in [27], the reliability of MEAM for description of bcc Ti properties is established by comparing CMD vs. experimental phonon dispersion curves [33,49] and total diffusion coefficients $D_{tot}$ [9,34-36], see [30]. The temperature-dependent effective potential (TDEP) method [37-39] is used to calculate the phonon spectra of bcc Ti (5488-atom supercell) by fitting second order force constants to displacement/force data sets extracted from CMD simulations. Videos [30] and snapshots are produced using VMD [40].

Figure 1 illustrates concerted migration events in defect-free bcc Ti recorded during AIMD at 1800 K. The figure is a superposition of snapshots in which atom-color variations are indicative of time progression during 9 ps. Two separate processes are identified: (*i*) direct exchange of two neighbors, which lasts for <1 ps; (*ii*) concerted migration of 22 neighbors along a tangled closed-loop trajectory, with duration of ≈9 ps. The path intersects supercell boundaries and follows both <111> (nearest-neighbor) and <100> (next-nearest-neighbor) directions.

CMD simulations carried out for 5488-atom defect-free bcc Ti supercells at temperatures between 0.73 and 0.97·$T_m$ complement the information provided by AIMD simulations at 1800 K (0.93·$T_m$). In agreement with AIMD observations, CMD runs show that the reaction pathways for concerted Ti migration include Ti-pair exchange, 4-atom ring rotation, or synchronized migration of dozens of neighbors (see section S5 with additional illustration [30]). Extended concerted diffusion events are initiated with a Ti atom that, by pushing a neighbor, triggers a chain of migration reactions. The reaction duration scales with the number of atoms involved; the ideal lattice is broken and reformed within ~$10^{-13}$–$10^{-12}$ s (Ti-pair exchange) up to ≈1 ns (collective motion of dozens of atoms). Although concerted migration phenomena as, e.g., the cyclic motion of few atoms, are believed to control diffusion in liquid and superheated metals [15,41], this type of mass transport processes has not been previously reported – except for bcc Fe at inner-core conditions [18] – in elemental solid crystals below the melting point.

The activation energy and attempt frequency of concerted migration evaluated via Arrhenius linear regression of CMD rates in defect-free cells [30] are $E_{aC}$=3.1±0.2 eV and $\nu_C$=1.0($\times 5.2^{\pm 1}$)$\times 10^{15}$ s$^{-1}$atom$^{-1}$ (inset Fig. 2a). The probability for one lattice atom to trigger collective motion at 1800 K (2.1$\times 10^{6\pm 0.3}$ s$^{-1}$atom$^{-1}$) is reasonably consistent with the estimate (≈$10^{8\pm 1.5}$ s$^{-1}$atom$^{-1}$ [42]) of a single AIMD run (250 atoms) at the same temperature. Although concerted migration rates exhibit Arrhenius-like behavior, the slope of corresponding self-diffusion coefficients $D_C(T)$ significantly varies with the inverse temperature (orange squares, Fig. 2a). For $T$ increasing from ≈0.73 to 0.97·$T_m$, the log[$D_C(T)$] vs. $T^{-1}$ curve changes concavity (upward→downward) at ≈0.85·$T_m$. Simulations performed for supercells that contain one vacancy show that concerted diffusion also enhances vacancy transport; collective motion of a <111> atomic-chain formed of *n* nearest-neighbors produces a net vacancy translation by *n* lattice positions.

For migration of individual vacancies, we obtain activation energies $E_{aV}$=0.19±0.02 eV and prefactors $D_{0V} \cdot c_V^{-1}$=6.2(×1.2$^{\pm 1}$)×10$^{-8}$ m$^2$s$^{-1}$ (red hexagons, Fig. 2b), where $c_V$≈0.2% is kept constant with $T$. The contribution of vacancy migration to the total diffusivity becomes progressively less important as $T$ increases. This is qualitatively demonstrated in Fig. 2a, which shows that the difference between self-diffusion coefficients in 5488-site supercells with (green circles) and without (orange squares) defects becomes vanishingly small near $T_m$.

The highly correlated motion of several Ti neighbors over a relatively flat energy landscape (≈0.2 eV barrier for vacancy migration) resembles the liquid-like diffusion properties characteristic of superionics (migration energies ≈0.2–0.3 eV) [43]. Also the trend in log[$D_C(T)$] vs. $T^{-1}$ determined for defect-free bcc Ti (orange squares, Fig. 2a) is qualitatively similar to that reported for bulk anion diffusivities in the type-II superionics $CaF_2$ and $UO_2$ (note change in curvature for temperatures approaching superionic transitions in figure 7 of [12]).

Superionic conductors are classified as type-I or type-II, respectively, depending on whether, or not, the transition to the disordered superionic state is accompanied by a structural transformation [44]. The diffusion behavior in bcc Ti is reminiscent of the one observed in type-II superionics in that concerted migration becomes dominant at $T$>0.8·$T_m$ [11] and no structural transition is associated with the onset of high diffusivity. At variance with superionics, where liquid-like diffusion is present in one sublattice while the remaining system stays ordered, bcc Ti shows localized liquid-like diffusion while also acting as solid host crystal.

The spontaneous formation of Frenkel defects (vacancy/self-interstitial pairs) is considered of central importance for promoting liquid-like diffusion in type-II superionics [44] and superheated crystals [15]. In particular Zhang *et al.* [15] noted that interstitial defects serve as initiators of string-like diffusion in superheated Ni and that the concerted process itself can be viewed as the propagation of interstitials along the string. It may be possible to make a similar interpretation of the concerted migration in bcc Ti, although analyzing this in terms of energy-quenched CMD snapshots, as in [15], is not possible here due to the dynamical instability of the bcc structure at 0 K [27]. We can, however, make some observations regarding the energetics of defect formation. Calculated vacancy equilibrium concentrations follow an Arrhenius trend within the investigated temperature range (Fig. 2S [30]). Linear regression of $\log[c_V(T)]$ and $\log[c_{SI}(T)]$ vs. $T^{-1}$ (Table I) yields free-energies of formation $G_V^f=1.00\pm0.05$ eV and $G_{SI}^f\approx3.5$ eV. The fact that $G_{SI}^f+G_V^f\approx4.5$ eV is much greater than the activation energy $E_{aC}$ (=3.1±0.2 eV) indicates that concerted diffusion in bcc Ti is more complex than the *static* formation of a Frenkel pair followed by rapid vacancy/self-interstitial migration and recombination.

Our observation of concerted migration may resolve several longstanding questions concerning mass transport and phonon properties of bcc Ti and, possibly, bcc Zr and Hf. Experimental studies attributed the anomalies in mass transport properties of Group-IVB transition metals to features in their vibrational spectra [7]. Special attention was dedicated to analyze the effects of soft phonon modes. Neutron scattering measurements indicated that the primary migration mechanism in bcc Ti is a simple <111> nearest-neighbor vacancy jump [45]. Rapid vacancy migration is enabled by soft 2/3<111> longitudinal modes (ω mode, Fig. 3a), a characteristic of bcc-structure Group-IVB metals [33,46-48], which assist atomic displacements along <111> directions [48].

The ω mode brings closer pairs of adjacent (111) lattice planes while leaving unaltered positions in every third (111) atomic layer (Fig. 3b,c). ω vibrations also correspond to parallel adjacent <111> atomic rows (strings) sliding one onto another maintaining unvaried interatomic spacing within each string (Fig. 3b,c). The ω-mode softness in Group-IVB bcc metals originates from weak inter-string forces (due, in turn, to d-electron screening) and electron accumulation along individual strings [7,49]. The observation of long <111> atomic-chains rapidly gliding against each other, Fig. 1, suggests that concerted migration is a consequence of soft ω modes. ω-like environments act as transition states for collective atomic motions characterized by continuous bcc→ω→bcc local structural changes (Fig. 3c).

The upward curvature in self-diffusivities $\log[D(T)]$ vs. $T^{-1}$, observed upon cooling bcc Ti toward the martensitic transformation temperature, has been correlated with the progressive softening of transversal $T_1$ ½<110> phonon modes, which lowers the vacancy migration enthalpy by causing distortion of the saddle-point configuration [7]. We note, however, that the curvature in log[$D_V(T)$] vs. $T^{-1}$ {$D_V(T) \propto c_V(T) \cdot \Gamma_V(T)$; $\Gamma_V(T)$ are vacancy jump frequencies} may also arise from non-Arrhenius trends in vacancy equilibrium concentrations $c_V(T)$ [50]. While experimental studies focused on understanding the breakdown of the Arrhenius law for $T$ approaching martensitic transitions, our results provide clarification for trends and magnitudes of Ti self-diffusivities at temperatures above $0.7 \cdot T_m$.

Present CMD results rule out both scenarios of $c_V(T)$ (Fig. S2 [30]) and $\Gamma_V(T)$ (red hexagons, Fig. 2b) displaying non-Arrhenius behavior for $T>0.7 \cdot T_m$. Figure S3 [30] demonstrates that the total diffusion coefficients $D_{tot}(T)$ evaluated via simulations of Ti(001) slabs are in agreement with experimental findings [9,34-36]. A comparison of $D_{tot}(T)$ values with CMD simulation results obtained for smaller simulation boxes (details given above) reveals that the total Ti self-diffusivity is essentially due to vacancy migration for $T<0.8 \cdot T_m$ and concerted migration (irrespective of defect concentrations) for $T>0.8 \cdot T_m$. This is summarized as $D_{tot}(T) \approx D_V(T)+[1-c_V(T)] \cdot D_C(T)$ [51], where $D_V(T)=c_V(T) \cdot D_V^{Cv\text{-indep.}}(T)$ and $D_V^{Cv\text{-indep.}}(T)$ is the diffusivity of an individual vacancy [30]. Concerted migration provides important contributions to mass transport at elevated $T$, which explains strong upward bending of log[$D_{tot}$] vs. $T^{-1}$ experimentally observed for temperatures approaching $T_m$ (inset in Fig. 2b and Fig. S3 [30]). This is demonstrated, Fig. 2b, by comparing the diffusion coefficients $D(T) \cdot c_V^{-1}$ for large vs. small supercells (both contain one vacancy, i.e., $c_V(T)$ is constant). In smaller supercells, only vacancy migration contributes to $D(T)$ at all temperatures. Vacancy migration is the dominant transport mechanism for large cells at relatively low temperatures (note hexagons/pentagons overlap for $T_m/T>1.2$, Fig. 2b), while the frequent occurrence of concerted migration in defect-free lattice regions for $T_m/T<1.2$ produces an upward curvature in log[$D(T) \cdot c_V^{-1}$] vs. $T_m/T$.

In addition to providing fundamental understanding for Ti mass transport properties, the occurrence of liquid-like diffusion clarifies puzzling features observed in bcc Ti phonon spectra [33,48]. CMD simulations well reproduce the phonon frequencies determined via neutron scattering measurements on bcc Ti, Fig. 3a. In agreement with experiments, CMD phonon dispersions exhibit no appreciable variation in ω-mode frequencies and reproduce the softening of $T_1$ ½<110> modes upon lowering T toward martensitic transition temperatures (Fig. S4 [30]). In Fig. 3a, one can note that the experiments detect a wide range of phonon energies (these reach vanishingly small values) in correspondence of the ω mode [52]. We argue below that such peculiarity is due to the occurrence of concerted migration.

The experimental study in [45] discussed possible origins for zero-frequency ω modes detected by neutron scattering in bcc Ti. The authors [45] excluded presence of static ω-structure embryos while noted, instead, that (*i*) ω phonons have lifetimes of ~$10^{-13}$ s and (*ii*) no coherent elastic intensity is detected for ħω=0 at **q**=2/3<111>. In this regard, Petry *et al.* [33] stated that a loss in real-space correlation combined with short phonon lifetimes is indicative of a liquid-like behavior in a localized region (near the ω point) of reciprocal space. Since atomic diffusion corresponds to zero-frequency vibrational modes of the system [53] and that string-like concerted migration closely relates with atomic displacements produced by ω phonons (Fig. 3b), we suggest that concerted diffusion is the atomistic manifestation of the peculiar feature observed in bcc Ti phonon spectra.

We can make yet another analogy between the behavior of bcc Ti and type-II superionics. A particular phonon mode of $B_{1u}$ symmetry appears to be connected to the superionic behavior in type-II superionic transitions [54-58]. The atomic displacements in this mode are closely connected to diffusion processes in fluorite crystal structures, analogous to the case of ω-mode displacements/concerted-migration in bcc Ti. Similar to the Ti ω-mode, the $B_{1u}$ mode is not critically soft in TDEP descriptions [58], but is so in the quasiharmonic case.

To summarize, atomistic simulations show that highly concerted string-like diffusion mechanisms are active in bcc Ti at temperatures below melting. Such diffusion processes have been previously observed for exotic materials, like superionic conductors and superheated crystals, or for iron under extreme temperatures and pressures. We propose that this diffusion mechanism may address fundamental questions concerning anomalies in mass transport and phonon properties of bcc Ti. Given the similarities in the properties of Group-IVB metals, concerted migration is expected to be operative in bcc Zr [46] and Hf [47].

**Acknowledgements**
All simulations were carried out using the resources provided by the Swedish National Infrastructure for Computing (SNIC), on the Gamma and Triolith Clusters located at the National Supercomputer Centre (NSC) in Linköping, on the Beskow cluster located at the Center for High Performance Computing (PDC) in Stockholm, and on the Kebnekaise cluster located at the High Performance Computing Center North (HPC2N) in Umeå, Sweden. D.G.S. gratefully acknowledges financial support from the Olle Engkvist Foundation. Financial support from the Swedish Research Council (VR) through Grant No. 2015-04391 and No. 2014-4750, the Swedish Government Strategic Research Area in Materials Science on Functional Materials at Linköping University (Faculty Grant SFO-Mat-LiU No. 2009-00971), and the VINN Excellence Center Functional Nanoscale Materials (FunMat-2) Grant 2016–05156 is gratefully acknowledged.

**References**

[1] T. Amino, K. Arakawa, H. Mori, Activation energy for long-range migration of self-interstitial atoms in tungsten obtained by direct measurement of radiation-induced point-defect clusters, Philosophical Magazine Letters **91**, 86 (2011).
[2] K. Urban, A. Seeger, Radiation-induced diffusion of point-defects during low-temperature electron irradiation, Philosophical Magazine **30**, 1395 (1974).
[3] H.R. Paneth, The Mechanism of Self-Diffusion in Alkali Metals, Physical Review **80**, 708 (1950).
[4] P.M. Derlet, D. Nguyen-Manh, S.L. Dudarev, Multiscale modeling of crowdion and vacancy defects in body-centered-cubic transition metals, Physical Review B **76**, 054107 (2007).
[5] P.-W. Ma, S.L. Dudarev, Universality of point defect structure in body-centered cubic metals, Physical Review Materials **3**, 013605 (2019).
[6] Z. C., Ring diffusion in metals, Acta Crystallographica **3**, 346 (1950).
[7] C. Herzig, The correlation between diffusion behavior and phonon softening in bcc metals, Berichte Der Bunsen-Gesellschaft Physical Chemistry **93**, 1247 (1989).
[8] C. Herzig, U. Kohler, S.V. Divinski, Tracer diffusion and mechanism of non-Arrhenius diffusion behavior of Zr and Nb in body-centered cubic Zr-Nb alloys, Journal of Applied Physics **85**, 8119 (1999).
[9] U. Kohler, C. Herzig, On the anomalous self-diffusion in bcc Titanium, Physica Status Solidi B **144**, 243 (1987).
[10] G.B. Gibbs, D. Graham, D.H. Tomlin, Diffusion in titanium and titanium—niobium alloys, Philosophical Magazine **8**, 1269 (1963).
[11] S. Hull, Superionics: crystal structures and conduction processes, Reports on Progress in Physics **67**, 1233 (2004).
[12] A. Annamareddy, J. Eapen, Low Dimensional String-like Relaxation Underpins Superionic Conduction in Fluorites and Related Structures, Scientific Reports **7**, 44149 (2017).
[13] V.A. Annamareddy, P.K. Nandi, X. Mei, J. Eapen, Waxing and waning of dynamical heterogeneity in the superionic state, Physical Review E **89**, 010301(R) (2014).
[14] A. Gray-Weale, P.A. Madden, Dynamical arrest in superionic crystals and supercooled liquids, Journal of Physical Chemistry B **108**, 6624 (2004).
[15] H. Zhang, M. Khalkhali, Q. Liu, J.F. Douglas, String-like cooperative motion in homogeneous melting, Journal of Chemical Physics **138**, 12A538 (2013).
[16] C. Donati, S.C. Glotzer, P.H. Poole, W. Kob, S.J. Plimpton, Spatial correlations of mobility and immobility in a glass-forming Lennard-Jones liquid, Physical Review E **60**, 3107 (1999).
[17] X.P. Tang, U. Geyer, R. Busch, W.L. Johnson, Y. Wu, Diffusion mechanisms in metallic supercooled liquids and glasses, Nature **402**, 160 (1999).
[18] A.B. Belonoshko, T. Lukinov, J. Fu, J. Zhao, S. Davis, S.I. Simak, Stabilization of body-centred cubic iron under inner-core conditions, Nature Geoscience **10**, 312 (2017).
[19] R. Car, M. Parrinello, Unified approach for molecular dynamics and density-functional theory, Physical Review Letters **55**, 2471 (1985).
[20] G. Kresse, J. Furthmuller, Efficient iterative schemes for *ab initio* total-energy calculations using a plane-wave basis set, Physical Review B **54**, 11169 (1996).
[21] P.E. Blöchl, Projector augmented-wave method, Physical Review B **50**, 17953 (1994).
[22] J.P. Perdew, K. Burke, M. Ernzerhof, Generalized gradient approximation made simple, Physical Review Letters **77**, 3865 (1996).
[23] Z.-G. Mei, S.-L. Shang, Y. Wang, Z.-K. Liu, Density-functional study of the thermodynamic properties and the pressure-temperature phase diagram of Ti, Physical Review B **80**, 104116 (2009).
[24] X.G. Lu, M. Selleby, B. Sundman, Assessments of molar volume and thermal expansion for selected bcc, fcc and hcp metallic elements, Calphad **29**, 68 (2005), and references therein.
[25] S. Plimpton, Fast parallel algorithms for short-range molecular dynamics, Journal of Computational Physics **117**, 1 (1995).

[26] B.J. Lee, M.I. Baskes, Second nearest-neighbor modified embedded-atom-method potential, Physical Review B **62**, 8564 (2000).
[27] W.-S. Ko, B. Grabowski, J. Neugebauer, Development and application of a Ni-Ti interatomic potential with high predictive accuracy of the martensitic phase transition, Physical Review B **92**, 134107 (2015).
[28] M. Parrinello, A. Rahman, Polymorphic transitions in single crystals – A new molecular dynamics method, Journal of Applied Physics **52**, 7182 (1981).
[29] Experimental and CMD results are compared as a function of normalized temperatures $T_m/T$: the Ti melting point is ≈1940 K; MEAM estimates $T_m$=1651 K [27].
[30] See supplemental information.
[31] Below, $D_{tot}$ is well approximated as sum of vacancy $D_V(T)$ and concerted migration $D_C(T)$ diffusion coefficients which, as described above, are separately determined via CMD simulations of 432- and 5488-bcc-site supercells.
[32] M.I. Mendelev, Y. Mishin, Molecular dynamics study of self-diffusion in bcc Fe, Physical review B **80**, 144111 (2009).
[33] W. Petry, A. Heiming, J. Trampenau, M. Alba, C. Herzig, H.R. Schober, G. Vogl, Phonon dispersion of the bcc phase of Group-IV metals. 1. bcc Titanium, Physical Review B **43**, 10933 (1991).
[34] J.F. Murdock, T.S. Lundy, E.E. Stansbury, Diffusion of $Ti^{44}$ and $V^{48}$ in titanium, Acta Metallurgica **12**, 1033 (1964).
[35] N.E. Walsöe De Reca, C.M. Libanati, Self diffusion in beta Titanium and beta Hafnium, Acta Metallurgica **16**, 1297 (1968).
[36] A.E. Pontau, D. Lazarus, Diffusion of titanium and niobium in bcc Ti-Nb alloys, Physical Review B **19**, 4027 (1979).
[37] O. Hellman, I.A. Abrikosov, S.I. Simak, Lattice dynamics of anharmonic solids from first principles, Physical Review B **84**, 180301(R) (2011).
[38] O. Hellman, P. Steneteg, I.A. Abrikosov, S.I. Simak, Temperature dependent effective potential method for accurate free energy calculations of solids, Physical Review B **87**, 104111 (2013).
[39] O. Hellman, I.A. Abrikosov, Temperature-dependent effective third-order interatomic force constants from first principles, Physical Review B **88**, 144301 (2013).
[40] W. Humphrey, A. Dalke, K. Schulten, VMD: Visual molecular dynamics, Journal of Molecular Graphics & Modelling **14**, 33 (1996).
[41] I.V. Belova, T. Ahmed, U. Sarder, A.V. Evteev, E.V. Levchenko, G.E. Murch, The Manning factor for direct exchange and ring diffusion mechanisms, Philosophical Magazine **97**, 230 (2017).
[42] The uncertainty on the AIMD rate is assessed from the scatter in $\Gamma_C(T)$ recorded from several CMD simulations at a given $T$.
[43] X. He, Y. Zhu, Y. Mo, Origin of fast ion diffusion in super-ionic conductors, Nature Communications **8**, 15893 (2017).
[44] A.K. Sagotra, D. Errandonea, C. Cazorla, Mechanocaloric effects in superionic thin films from atomistic simulations, Nature Communications **8**, 963 (2017).
[45] W. Petry, T. Flottmann, A. Heiming, J. Trampenau, M. Alba, G. Vogl, Atomistic study of anomalous self-diffusion in bcc β-Titanium, Physical Review Letters **61**, 722 (1988).
[46] A. Heiming, W. Petry, J. Trampenau, M. Alba, C. Herzig, H.R. Schober, G. Vogl, Phonon dispersion of the bcc phase of Group-IV metals. 2. bcc Zirconium, Physical Review B **43**, 10948 (1991).
[47] J. Trampenau, A. Heiming, W. Petry, M. Alba, C. Herzig, W. Miekeley, H.R. Schober, Phonon dispersion of the bcc phase of Group-IV metals. 3. bcc Hafnium, Physical Review B **43**, 10963 (1991).
[48] W. Petry, A. Heiming, J. Trampenau, M. Alba, G. Vogl, Strong phonon softening in the bcc phase of Titanium, Physica B **156**, 56 (1989).

[49] K.M. Ho, C.L. Fu, B.N. Harmon, Microscopic analysis of interatomic forces in transition-metals with lattice distortions, Physical Review B **28**, 6687 (1983).

[50] A. Glensk, B. Grabowski, T. Hickel, J. Neugebauer, Breakdown of the Arrhenius Law in Describing Vacancy Formation Energies: The Importance of Local Anharmonicity Revealed by *Ab initio* Thermodynamics, Physical Review X **4**, 011018 (2014).

[51] $D_{tot}(T)$ slightly larger than $D_V(T)+[1-c_V(T)] \cdot D_C(T)$ due to defect/defect interactions and $Ti_{SI}$ migration in slab simulations.

[52] TDEP is unsuited to reproduce that feature: it requires atomic trajectories to oscillate near fixed lattice positions.

[53] M.P. Desjarlais, First-principles calculation of entropy for liquid metals, Physical Review E **88**, 062145 (2013).

[54] L.L. Boyer, Nature of melting and superionicity in alkali and alkaline-earth halides, Physical Review Letters **45**, 1858 (1980).

[55] J. Buckeridge, D.O. Scanlon, A. Walsh, C.R.A. Catlow, A.A. Sokol, Dynamical response and instability in ceria under lattice expansion, Physical Review B **87**, 214304 (2013).

[56] M.K. Gupta, P. Goel, R. Mittal, N. Choudhury, S.L. Chaplot, Phonon instability and mechanism of superionic conduction in $Li_2O$, Physical Review B **85**, 184304 (2012).

[57] J.R. Nelson, R.J. Needs, C.J. Pickard, High-pressure $CaF_2$ revisited: A new high-temperature phase and the role of phonons in the search for superionic conductivity, Physical Review B **98**, 224105 (2018).

[58] J. Klarbring, N.V. Skorodumova, S.I. Simak, Finite-temperature lattice dynamics and superionic transition in ceria from first principles, Physical Review B **97**, 104309 (2018).

**Tables/Figures**

| $T_m/T$ | $c_V$ | $c_{SI}$ |
|---|---|---|
| 1.38 | $(4.0\pm0.1)\cdot10^{-5}$ | – |
| 1.27 | $(6.7\pm0.1)\cdot10^{-5}$ | – |
| 1.18 | $(1.3\pm0.1)\cdot10^{-4}$ | $(1.3\pm0.1)\cdot10^{-5}$ |
| 1.10 | $(2.0\pm0.1)\cdot10^{-4}$ | $(6.7\pm0.1)\cdot10^{-5}$ |

**Table I.** Equilibrium concentration of vacancies and self-interstitials in bcc Ti estimated via CMD.

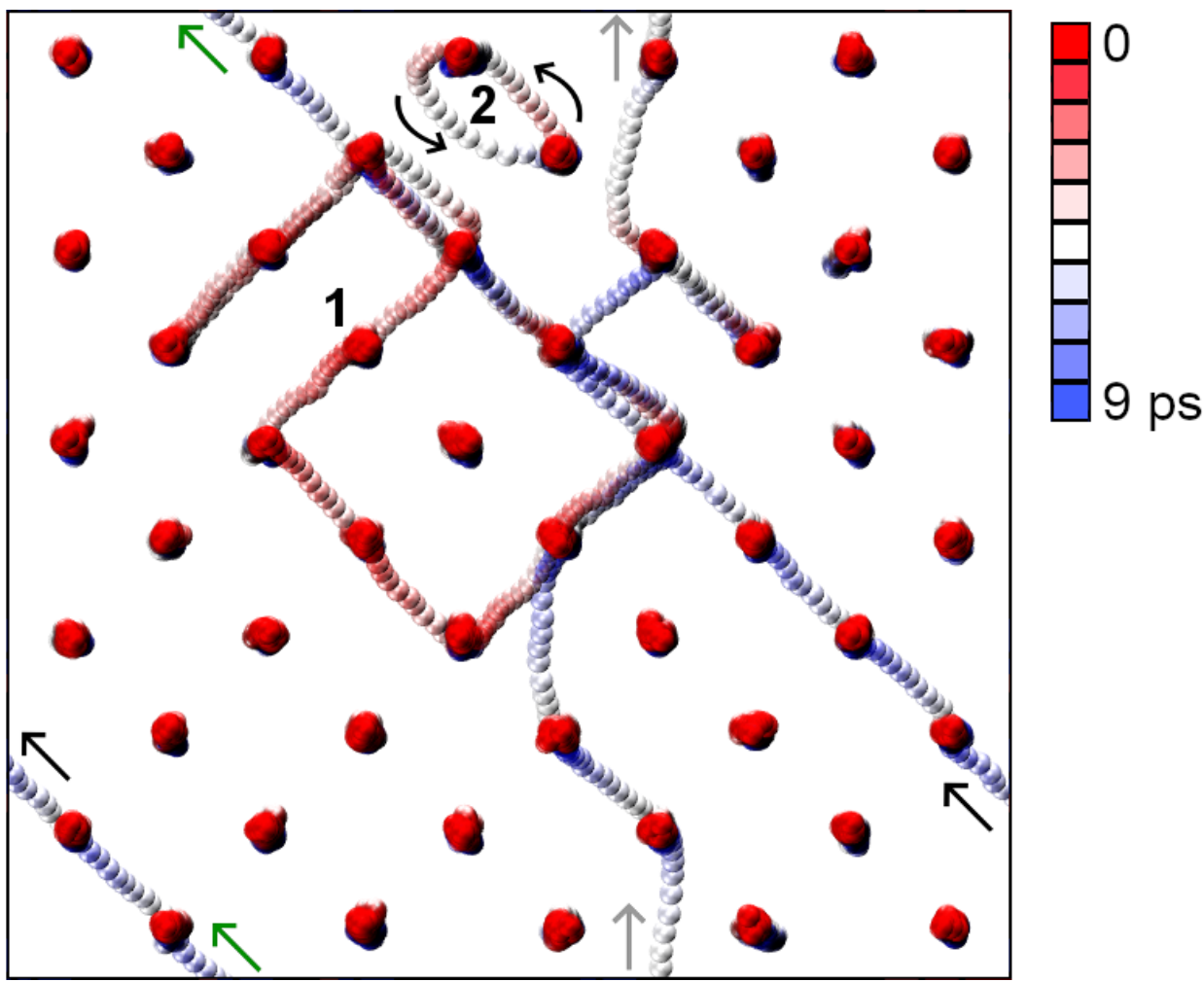

**Fig. 1.** Orthographic (001)-plane view of overlapped AIMD snapshots taken during concerted migration in defect-free bcc Ti at 1800 K. In each snapshot, atoms are colored according to time progression during ≈9 ps (see legend). Identified events are labeled as "**1**" (concerted migration of 22 atoms) and "**2**" (Ti-pair exchange). The *complex* mechanism **1** entails that a tangled closed-loop atomic chain – with <111> and <001> chain-segments linking nearest-neighbor and next-nearest-neighbors – glides in a synchronized cyclic fashion within the solid host crystal. During the reaction, each chain-atom replaces the next on the chain sequence. Straight colored arrows facilitate following atomic displacements across supercell boundaries. For clarity, trajectory-smoothing over 1000 steps is used to reduce the vibrational noise. Simulation videos and additional description of concerted diffusion mechanisms (section S5) are provided in supplemental files [30].

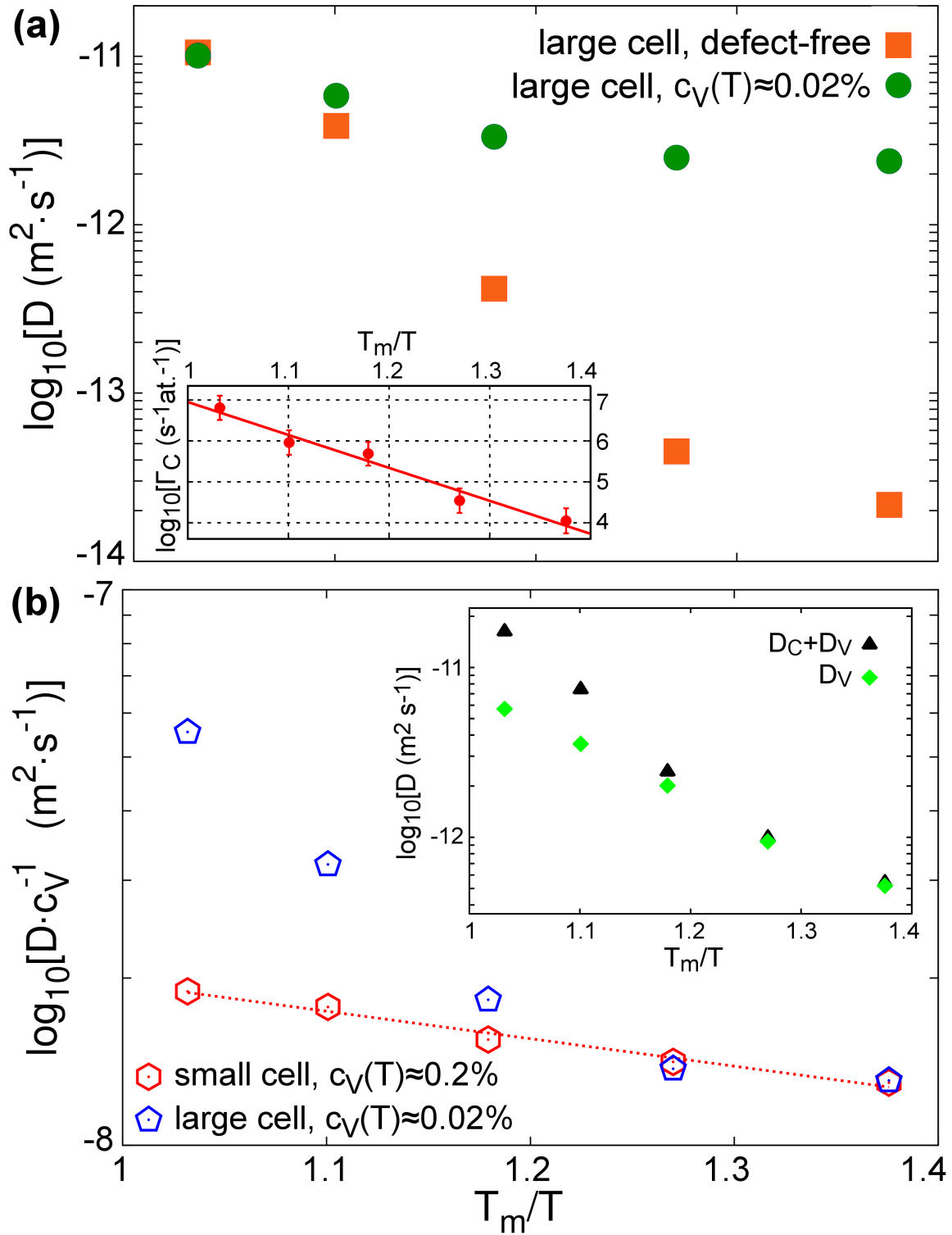

**Fig. 2.** Temperature dependence of Ti self-diffusion due to vacancy and/or collective-atomic-motion determined via CMD. The labels *large*(*small*) indicate results of supercells containing 5488(432) bcc lattice sites. **(a)** The supercells contain one vacancy (green circles) or no defects (orange squares=$D_C$). The inset illustrates the temperature-dependence of concerted migration rates $\Gamma_C$. **(b)** Vacancy-concentration-normalized diffusion coefficients as a function of supercell size; both large and small supercells contain one vacancy. In smaller cells (red hexagons), diffusion is exclusively due to vacancy migration (no diffusion in defect-free regions). Concerted migration in defect-free regions of larger cells (blue pentagons) yields significant diffusivities for $T_m/T<1.2$. The inset shows results for $D_V$ (diffusivity of an individual vacancy multiplied by $c_V(T)$) vs. total self-diffusivities (approximated as $D_C+D_V$) [30].

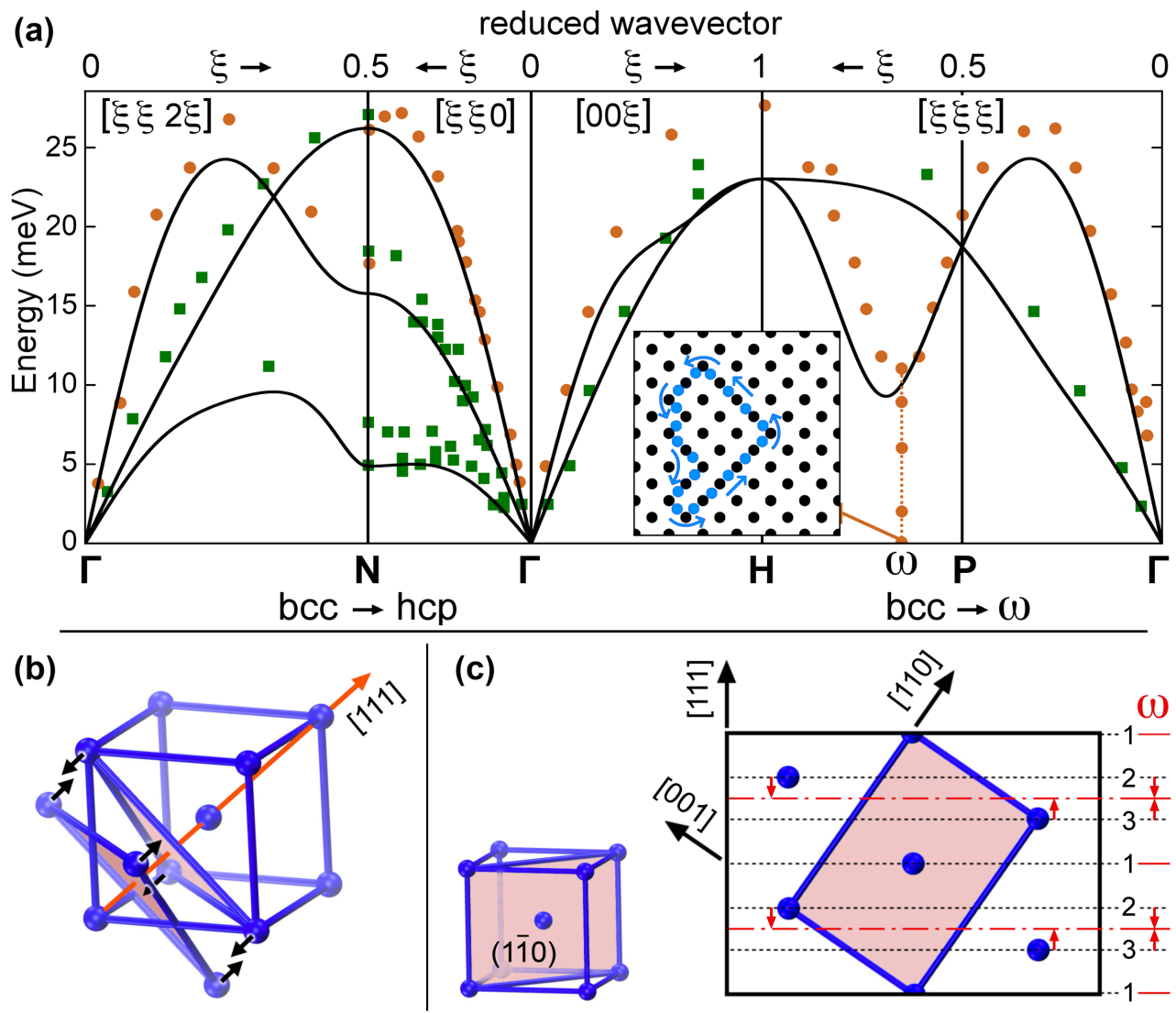

**Fig. 3. (a)** Titanium phonon dispersion evaluated via CMD (solid lines) vs. inelastic neutron scattering data at ≈1300 K. Experimental longitudinal/transversal-mode frequencies are marked by orange/green circles/squares. Note that experiments [33,48] report different frequencies for T₁ modes along Γ→N paths. The inset in **(a)** is a schematic illustration of liquid-like diffusion, here suggested to underlie anomalously-low ω frequencies in bcc Ti. **(b,c)** Schematic representation of longitudinal 2/3[111] ω phonons responsible for bcc→ω phase transitions and concerted migration in titanium.

*Supplemental material*
**Superionic-like diffusion in an elemental crystal: bcc Titanium**
D.G. Sangiovanni,[1,2] J. Klarbring,[1] D. Smirnova,[2,3] N.V. Skripnyak,[1] D. Gambino,[1]
M. Mrovec,[2] S.I. Simak,[1] I.A. Abrikosov[1,4]

[1]Department of Physics, Chemistry and Biology (IFM) Linköping University,
SE-581 83, Linköping, Sweden

[2]ICAMS, Ruhr-Universität Bochum, D-44780 Bochum, Germany

[3]Joint Institute for High Temperatures, Russian Academy of Sciences, Moscow, Russia

[4]Materials Modeling and Development Laboratory, NUST "MISIS", 119049 Moscow, Russia

**S1. CMD evaluation of bcc Ti equilibrium volumes as a function of temperature.** The equilibrium volume of bcc Ti has been calculated using CMD simulations employing *NPT* sampling of the configurational space (timestep of 1 fs and simulation time of 1 ns for each thermostat temperature). *NPT* sampling is based on the Parrinello-Rahman barostat with pressure damping parameter = 1 ps and the Langevin thermostat with damping parameter = 0.1 ps. The supercell (14×14×14 conventional bcc cells = 5488 Ti atoms) maintains the bcc structure during the simulated time (≈1 ns). It should be stressed that the Ti melting point ($T_m$) predicted by MEAM is 1651 K [1], while the experimental value is ≈1940 K.

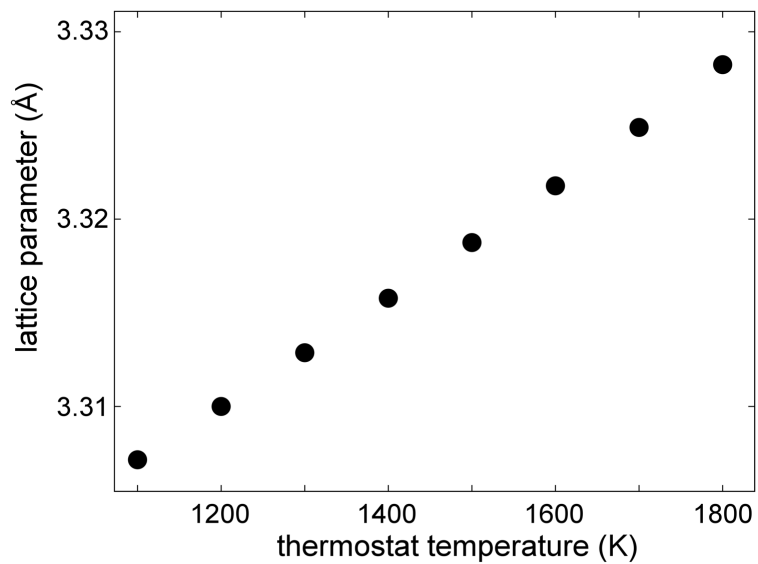

**Fig. S1.** Lattice parameter of bcc Ti determined for 5488-atom supercells via CMD simulations within the *NPT* ensemble.

**S2. Determination of migration frequencies and diffusivities with related activation energies, exponential prefactors, and their uncertainties.** Vacancy diffusivities $D_V$, as well as rates $\Gamma_C$ and diffusivities $D_C$ of concerted migration in defect-free lattices are determined by combining the results obtained for supercells formed of 6×6×6 (432 lattice sites) and 14×14×14 (5488 lattice sites) conventional bcc cells. CMD simulations employ *NVT* canonical sampling at temperatures *T*

ranging from 0.73 to 0.97·$T_m$ (Nosé-Hoover thermostat temperatures between 1200 and 1600 K). Prior to counting migration events or mean square displacements, the supercells (lattice parameters taken from *NPT* simulations, Fig. S1) are equilibrated for 50 ps.

Activation energies $E_a$, attempt frequencies $\nu$, and exponential prefactors $D_0$ are evaluated via Arrhenius linear regression of migration frequencies $\Gamma(T)$ or atomic diffusivities $D(T)$:

$$\ln[\Gamma(T)] = \ln(\nu) - \frac{E_a}{k_B}\frac{1}{T} \quad ; \quad \ln[D(T)] = \ln(D_0) - \frac{E_a}{k_B}\frac{1}{T}. \tag{S1}$$

The rates (per atoms) of concerted migration $\Gamma_C$ in defect-free bcc Ti are determined for 5488-atom supercells. A number ($n_r$) of ten, or more, independent runs $i$ is performed at each $T$. Thus, $\Gamma_C$ is calculated at each temperature as

$$\Gamma_C = \sum_i^{n_r} \delta_i \Bigg/ \sum_i^{n_r} t_i, \tag{S2}$$

where $\delta_i = 1$ (or 0) depending on whether concerted migration occurred (or not) during simulation $i$. In runs for which $\delta_i = 0$, $t_i$ is equal to the total simulation time. In cases for which $\delta_i = 1$, $t_i$ is the time required for an atom to initiate concerted migration. The results of Eq. (S2) correspond to calculating average jump rates via exponential-distribution statistics (see Appendix A in Ref. [2]). Uncertainties on $\Gamma_C$, $E_{aC}$, and $\nu_C$ values are calculated as described in Ref. [2]. The occurrence of concerted migration during CMD simulations is detected by tracing the displacement of atoms from ideal lattice positions, and verified by visualization of the simulation using VMD [3].

The diffusivity $D$ at a temperature $T$ in a supercell that contains $N$ atoms is computed by tracing the mean square displacement (MSD) of each atom $j$ during time $t$:

$$D(T) = \frac{\text{MSD}(t,T)}{6t} = \frac{1}{6t \cdot N} \sum_j^N \left[ \vec{r}_j(t,T) - \vec{r}_j(0,T) \right]^2. \tag{S3}$$

The equality in Eq. S3 is formally correct at the limit for $t \to \infty$. Hence, CMD simulation times are sufficiently long to ensure that the MSD evolves linearly with $t$. As discussed in the main text, the total Ti diffusivity $D_{tot}$ derives primarily from the combination of monovacancy and concerted migration. During ~1-ns-long CMD simulations, performed on relatively small simulation boxes (6×6×6 conventional bcc cells), vacancy migration occurs rapidly enough for MSD($t$) vs. $t$ to reach steady slopes, whereas concerted migration in the defect-free lattice region is never observed. This allows calculating the temperature dependence of vacancy diffusivities. Irrespective of temperature, the supercells used to evaluate the vacancy diffusivity have constant vacancy concentration $c_V = N^{-1}$ ($N = 432$, number of bcc lattice sites in the simulation box). The vacancy-concentration-independent diffusivity is calculated as

$$D_V^{c_V-\text{indep.}}(T) = \sum_{j=1}^{N-1} \frac{\left[\vec{r}_j(t,T) - \vec{r}_j(0,T)\right]^2}{6t}. \quad (S4)$$

The vacancy diffusivity $D_V(T)$ is obtained by multiplying $D_V^{c_V-\text{indep.}}(T)$ by the equilibrium vacancy concentration $c_V(T)$ evaluated via the slab-model approach described in Sec. S3:

$$D_V(T) = c_V(T) \cdot D_V^{c_V-\text{indep.}}(T). \quad (S5)$$

Atomic diffusivities due to concerted atomic migration $D_C$ are calculated by tracing the mean-square displacement of all atoms $N$ in larger (14×14×14 conventional bcc unit cells) defect-free supercells:

$$D_C(T) = \frac{1}{N} \cdot \sum_{j=1}^{N} \frac{\left[\vec{r}_j(t,T) - \vec{r}_j(0,T)\right]^2}{6t}. \quad (S6)$$

Neglecting the presence of self-interstitials, the total self-diffusion coefficient $D_{\text{tot}}$ is well approximated as

$$D(T) \approx D_V(T) + D_C(T) \cdot [1 - c_V(T)], \quad (S7)$$

for which $D_V(T)$ is obtained from Eq. (S5).

Rates and diffusivities of concerted migration in defect-free bcc Ti have also been tested during *NPT* sampling of the configurational space between 0.73 and 0.96·$T_m$. The results are consistent (within error bars) of *NVT* simulations.

**S3. Slab model.** Equilibrium concentrations of defects are calculated using the approach suggested in [4]. The model bcc Ti(001) surface slab has sizes of 40, 80, and 30·$a(T)$ in $x$, $y$, and $z$ Cartesian directions, respectively, for which $a(T)$, the bcc Ti lattice parameter at a temperature $T$, is determined via CMD/*NPT* simulations (Sec. S1). The free (001) surfaces are oriented along the $y$ direction. The slab contains 165600 atoms, including 91200 atoms in pre-surface areas (of width = 10·$a(T)$ on each side of the slab) and 74400 atoms in the *bulk* region used to evaluate equilibrium diffusivities and defect concentrations. The CMD simulations were performed in the *NVE* ensemble.

Initial simulation temperatures were set to ×2 desired target temperatures. During the simulations, we ensured that the temperature remained approximately constant at $T$ = 1200, 1300, 1400, and 1500 K, which correspond to the range 0.73 – 0.91·$T_m$. Typical simulation time-lengths were ≈1.5 ns, using 0.5-fs timestep. Tracking the number of vacancies formed during the calculation (by detecting atoms with the highest potential energy), we find that the concentration of defects in bcc Ti remains stationary after a simulation time of ≈0.8 ns, so that their concentration can be considered as equilibrium. During the calculations, we observe that most vacancies (recorded in the bulk region) originally formed at the surface. However, part of vacancy formation/annihilation is related to the formation of Frenkel pairs in the bulk. Hence, a small amount of equilibrium self-interstitials is detected in bcc Ti at elevated temperatures (Fig. S2). We should stress that the defect equilibrium concentrations may be inaccurate. Although hcp-Ti surface energies are well reproduced by MEAM [1], (*i*) bcc Ti surface properties have not been tested, (*ii*) slab terminations other than the (001) could modify estimated $c_V(T)$ and $c_{SI}(T)$ values.

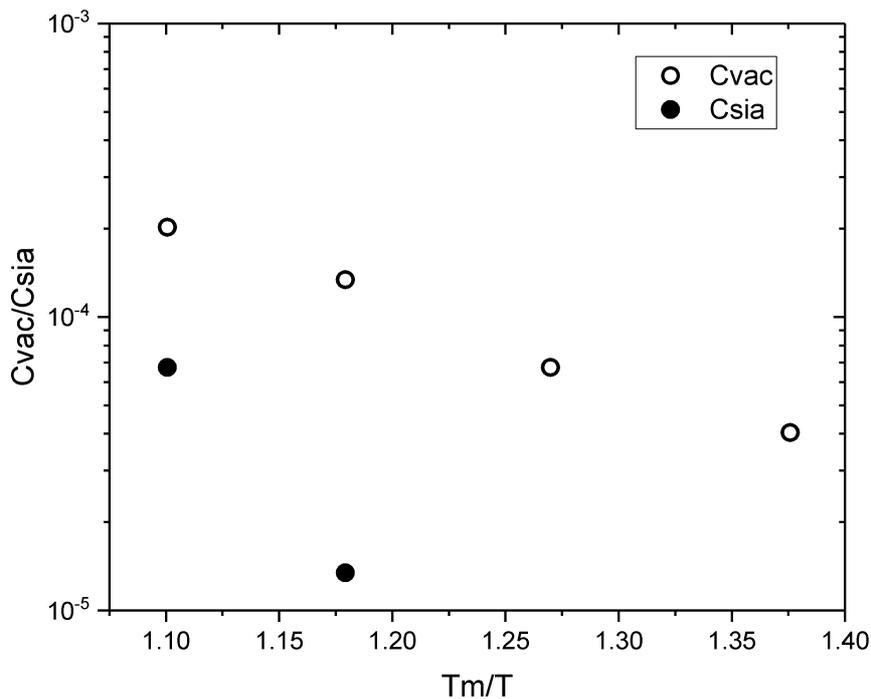

**Fig. S2.** Vacancy and self-interstitial equilibrium concentrations determined as a function of $T$.

As shown in Refs. [4, 5], the slab model can be also applied for the direct estimation of self-diffusion coefficients by tracing atomic mean-squared displacements in the bulk region (Eq. S3). The resulted self-diffusion coefficients ($D_{tot}$) include the contribution of all defects as well as of concerted migration in defect-free regions. Calculated $D_{tot}(T)$ values are in very good agreement with experimental values [6-9], see Fig. S3.

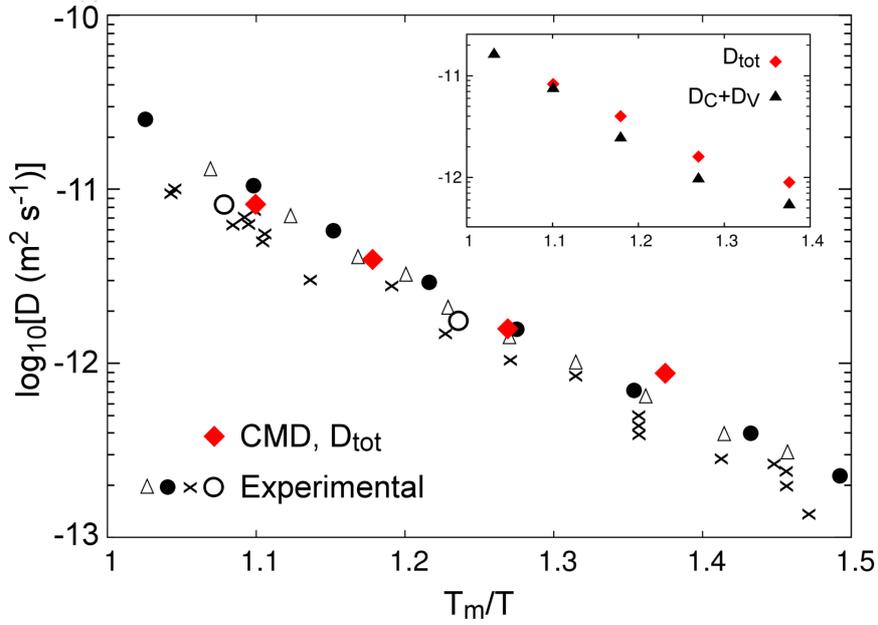

**Fig. S3.** Coefficient of self-diffusion in bcc Ti as predicted by CMD simulations in comparison with experimental results. Calculated error bars are smaller than the symbols. Experimental results are taken from (black filled circles [6], triangles [7], crosses [8], white circles [9]). The inset shows a comparison of calculated $D_{tot}$ (obtained using the slab model) with total diffusivities approximated as sum of vacancy and concerted migration contributions, as calculated via Eq. S7.

**S4. Calculated phonon spectra of bcc Ti.** The temperature-dependent effective potential (TDEP) method [10-12] is used to calculate the phonon spectra of bcc Ti (5488-atom supercell) by fitting second order force constants to displacement/force data sets directly extracted from CMD/*NVT* simulations. In Fig. S4, thermostat temperatures were set to 1100 and 1400 K. We note that, although the hcp→bcc transformation temperature is estimated around 1150 K, the bcc Ti phase is dynamically stable at 1100 K.

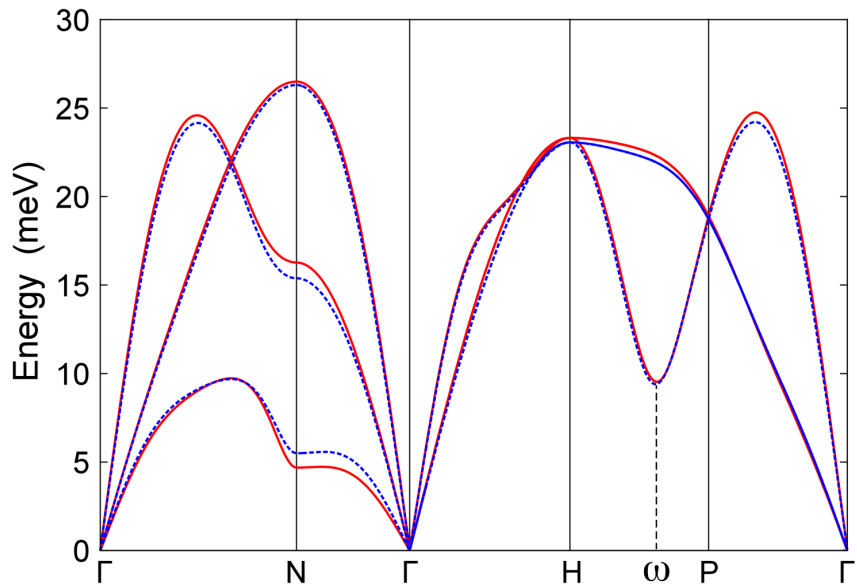

**Fig. S4.** Phonons dispersion calculated via CMD for bcc Ti. The thermostat is set at 1100 (red solid lines) and 1400 K (blue dashed line).

**S5. Additional description of concerted migration in defect-free bcc Ti.** CMD simulations confirm that concerted migration mechanisms, as observed in AIMD (see main text), are of relevance at temperatures far below the Ti melting point. To further facilitate the understanding of concerted migration, a complex mass transport process, we illustrate a particularly simple event recorded during CMD simulations. The reaction is initiated with five Ti atoms (labeled as #2 – #6, see color legend in Fig. S5) leaving their ideal lattice position at approximately the same time (Fig. S5a). The migration of Ti#2 promptly induces the neighbor atom #1 to move toward the vacated site (Fig. S5b). Similarly, the displacement of Ti#6 triggers a chain of atomic motions: Ti#7, Ti#8, and Ti#9 are sequentially pushed out from their ideal bcc sites by incoming Ti#6, Ti#7, and Ti#8 atoms, respectively (Fig. S5b). The snapshot in Fig. S5b is a representative example of simultaneous displacement of several atoms from ideal lattice positions. It is worth noting that the vacancy created by the migration of atom #1 attracts a neighbor Ti (see, on mid-left of panels b and c in Fig. S5, light-gray sphere that protrudes laterally out of its bcc site). However, the position formerly occupied by Ti#1 is rapidly taken by Ti#11 (Fig. S5c,d), which closes the concerted migration loop, thus reforming the ideal bcc lattice. Fig. S5e provides a schematic representation of atomic transport due to the concerted migration event.

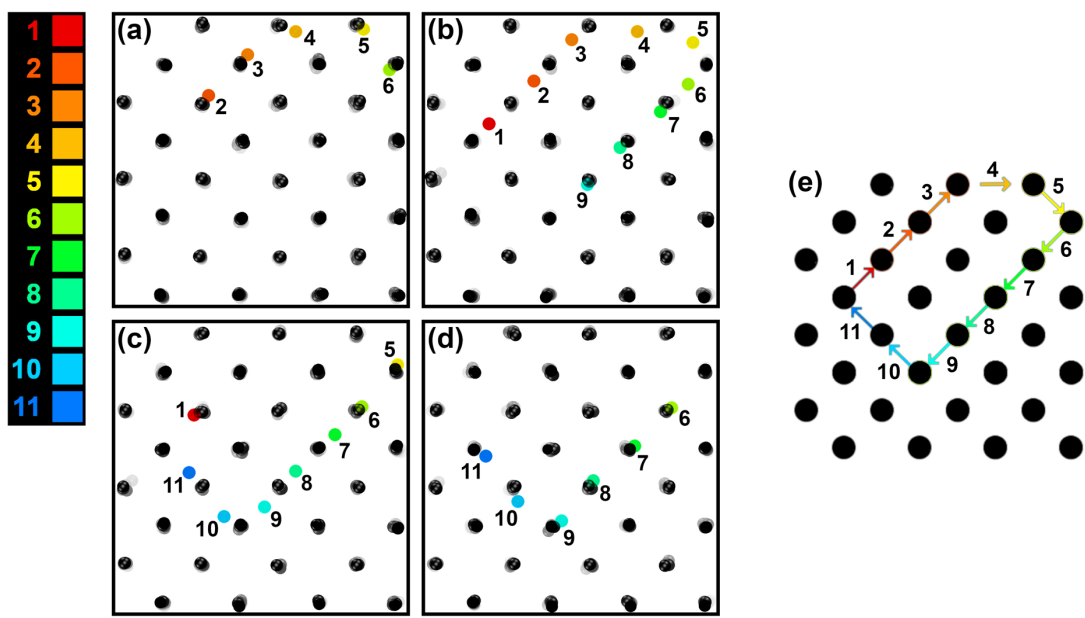

**Fig. S5.** Orthographic-view on the (001) crystallographic plane of defect-free bcc Ti during concerted migration, as described by CMD simulations at a thermostat temperature of 1600 K. The snapshots (a-d), taken at intervals of 0.8 ps, are focalized on a small region of the entire supercell (5488 atoms). Trajectory-smoothing over 1000 steps (1 ps) is used to reduce the vibrational noise. Depth-cueing allows distinguishing atoms in internal lattice layers (shaded gray spheres) from atoms in layers closer to the point of view (black spheres). In this example, string-like collective migration (a-d) entails a clockwise rotation of a closed-loop chain of neighbors. Panel (e) schematically summarizes <111> and <001> displacements (projected onto the (001) plane) for each of the 11 atoms that participate in the reaction.

**References (supplemental material)**

[1] W.-S. Ko, B. Grabowski, J. Neugebauer, Development and application of a Ni-Ti interatomic potential with high predictive accuracy of the martensitic phase transition, Physical Review B **92**, 134107, (2015).

[2] D. Gambino, D.G. Sangiovanni, B. Alling, I.A. Abrikosov, Nonequilibrium ab initio molecular dynamics determination of Ti monovacancy migration rates in B1 TiN, Physical Review B **96**, 104306 (2017).

[3] W. Humphrey, A. Dalke, K. Schulten, VMD: Visual molecular dynamics, Journal of Molecular Graphics & Modelling **14**, 33 (1996).

[4] M.I. Mendelev, Y. Mishin, Molecular dynamics study of self-diffusion in bcc Fe, Physical review B **80**, 144111 (2009).

[5] D.E. Smirnova, S.V. Starikov, I.S. Gordeev, Evaluation of the structure and properties for the high-temperature phase of zirconium from the atomistic simulations, Computational Materials Science **152**, 51 (2018).

[6] U. Kohler, C. Herzig, On the anomalous self-diffusion in bcc Titanium, Physica Status Solidi B **144**, 243 (1987).

[7] J.F. Murdock, T.S. Lundy, E.E. Stansbury, Diffusion of $Ti^{44}$ and $V^{48}$ in titanium, Acta Metallurgica **12**, 1033 (1964).

[8] N.E. Walsöe De Reca, C.M. Libanati, Self diffusion in beta Titanium and beta Hafnium, Acta Metallurgica **16**, 1297 (1968).

[9] A.E. Pontau, D. Lazarus, Diffusion of titanium and niobium in bcc Ti-Nb alloys, Physical Review B **19**, 4027 (1979).

[10] O. Hellman, I.A. Abrikosov, S.I. Simak, Lattice dynamics of anharmonic solids from first principles, Physical Review B **84**, 180301(R) (2011).

[11] O. Hellman, P. Steneteg, I.A. Abrikosov, S.I. Simak, Temperature dependent effective potential method for accurate free energy calculations of solids, Physical Review B **87**, 104111 (2013).

[12] O. Hellman, I.A. Abrikosov, Temperature-dependent effective third-order interatomic force constants from first principles, Physical Review B **88**, 144301 (2013).